\documentclass[aps,pra,reprint,amsmath,amssymb,superscriptaddress]{revtex4-2}

\usepackage[utf8]{inputenc}
\usepackage[T1]{fontenc}
\usepackage{graphicx}
\usepackage[colorlinks=true,urlcolor=blue,citecolor=blue,linkcolor=blue,bookmarks=false,pdfstartview={FitH}]{hyperref}

\usepackage{color,soul}

\begin{document}

\title{Strain-induced structural change and nearly-commensurate diffuse scattering in the model high-temperature superconductor HgBa$_2$CuO$_{4+\delta}$}

\author{Mai Ye}
\email{mai.ye@kit.edu}
\affiliation{Institute for Quantum Materials and Technologies, Karlsruhe Institute of Technology, Kaiserstr. 12, 76131 Karlsruhe, Germany}
\author{Wenshan Hong}
\affiliation{International Center for Quantum Materials, School of Physics, Peking University, 100871 Beijing, China}
\affiliation{Beijing National Laboratory for Condensed Matter Physics, Institute of Physics, Chinese Academy of Sciences, Beijing 100190, China}
\author{Tom Laurin Lacmann}
\altaffiliation[Present address: ]{Laboratory for Quantum Magnetism, Institute of Physics,
\'Ecole Polytechnique F\'ed\'erale de Lausanne, CH-1015 Lausanne, Switzerland}
\affiliation{Institute for Quantum Materials and Technologies, Karlsruhe Institute of Technology, Kaiserstr. 12, 76131 Karlsruhe, Germany}
\author{Mehdi Frachet}
\altaffiliation[Present address: ]{Institut N\'eel CNRS/UGA UPR2940, 25 Rue des Martyrs, 38042 Grenoble, France}
\affiliation{Institute for Quantum Materials and Technologies, Karlsruhe Institute of Technology, Kaiserstr. 12, 76131 Karlsruhe, Germany}
\author{Igor Vinograd}
\altaffiliation[Present address: ]{Laboratoire National des Champs Magn\'etigues Intenses, CNRS - Universit\'e Grenoble Alpes - Universit\'e Paul Sabatier - Institut National des Sciences Appliqu\'ees - European Magnetic Field Laboratory, 38042 Grenoble, France}
\affiliation{Institute for Quantum Materials and Technologies, Karlsruhe Institute of Technology, Kaiserstr. 12, 76131 Karlsruhe, Germany}
\author{Gaston Garbarino}
\affiliation{European Synchrotron Radiation Facility, BP 220, F-38043 Grenoble Cedex, France}
\author{Sofia-Michaela Souliou}
\affiliation{Institute for Quantum Materials and Technologies, Karlsruhe Institute of Technology, Kaiserstr. 12, 76131 Karlsruhe, Germany}
\author{Michael Merz}
\affiliation{Institute for Quantum Materials and Technologies, Karlsruhe Institute of Technology, Kaiserstr. 12, 76131 Karlsruhe, Germany}
\affiliation{Karlsruhe Nano Micro Facility (KNMFi), Karlsruhe Institute of Technology, Kaiserstr. 12, 76131 Karlsruhe, Germany}
\author{Rolf Heid}
\affiliation{Institute for Quantum Materials and Technologies, Karlsruhe Institute of Technology, Kaiserstr. 12, 76131 Karlsruhe, Germany}
\author{Amir-Abbas Haghighirad}
\affiliation{Institute for Quantum Materials and Technologies, Karlsruhe Institute of Technology, Kaiserstr. 12, 76131 Karlsruhe, Germany}
\author{Yuan Li}
\affiliation{International Center for Quantum Materials, School of Physics, Peking University, 100871 Beijing, China}
\affiliation{Beijing National Laboratory for Condensed Matter Physics, Institute of Physics, Chinese Academy of Sciences, Beijing 100190, China}
\author{Matthieu Le Tacon}
\email{matthieu.letacon@kit.edu}
\affiliation{Institute for Quantum Materials and Technologies, Karlsruhe Institute of Technology, Kaiserstr. 12, 76131 Karlsruhe, Germany}

\date{\today}

\begin{abstract}
We investigate the strain response of underdoped HgBa$_2$CuO$_{4+\delta}$ (Hg1201), by synchrotron X-ray diffraction and corresponding simulations of thermal diffuse scattering. The compression in the crystallographic \textit{a} direction leads to relatively small expansion in the \textit{b} and \textit{c} directions, with Poisson ratios $\nu_{ba}$\,=\,0.16 and $\nu_{ca}$\,=\,0.11, respectively. However, the Cu-O distance in the \textit{c} direction exhibits a notable 0.9\% increase at 1.1\% \textit{a}-axis compression. We further find strain-induced diffuse scattering which corresponds to a new type of two-dimensional charge correlation. Interestingly, this signal is insensitive to the onset of superconductivity and instead corresponds to a short-range, nearly commensurate modulation with a wave vector close to (0.5, 0, 0) and a correlation length of approximately four unit cells. It closely resembles the charge order theoretically predicted in the phase diagram of the spin-liquid model with resonating valence bonds on a square lattice.
\end{abstract}

\maketitle

\section{Introduction\label{sec:Intro}}
Strain tuning has emerged as a powerful tool for manipulating competing electronic orders and revealing new quantum phases in correlated systems: by modifying lattice symmetries, bond angles, and interatomic distances, strain can significantly alter the delicate balance between spin, charge, and superconducting orders, offering a controllable parameter to probe intertwined phenomena in the phase diagram~\cite{Tc2010,Hicks2014,Steppke2017,Tc2017,Y2018,Y2021,Y2022,Boyle2021,Najev2022,Keimer2022,Y2024,Lin2024,Sr2024}. Notably, uniaxial pressure has been shown to influence superconducting transition temperatures~\cite{Tc2010,Hicks2014,Steppke2017,Tc2017,Y2022} and stabilize charge-density-wave (CDW) phases that are otherwise suppressed under strain-free conditions~\cite{Y2018,Y2021,Y2024}.

A striking example is the emergence of long-range three-dimensional (3D) charge-density-wave (CDW) order in the double copper–oxygen-layer compound  YBa$_2$Cu$_3$O$_{y}$ (YBCO).  
This order was first revealed under high magnetic fields \cite{Mag2011,Mag2015}, and was later shown to be induced by suppressing the superconducting transition via uniaxial compression along the \(a\)-axis \cite{Y2018}.  
The observation of this otherwise hidden phase highlights a competing — or possibly intertwined — electronic order that coexists with, and is suppressed by, high-temperature superconductivity.

However, it remains an open question whether such strain-induced electronic ordering phenomena are universal across cuprate families or are limited to specific structural and electronic contexts. The special structure of YBCO, with its orthorhombic symmetry, CuO chains, and bilayer CuO$_2$ planes, introduces unique interlayer couplings and charge reservoirs that may influence the CDW formation under strain~\cite{Y2017,Y2024}. Thus, extending the investigation to structurally simpler cuprates is crucial for disentangling intrinsic effects from sample-specific characteristics.

\begin{figure}
\includegraphics[width=0.98\linewidth]{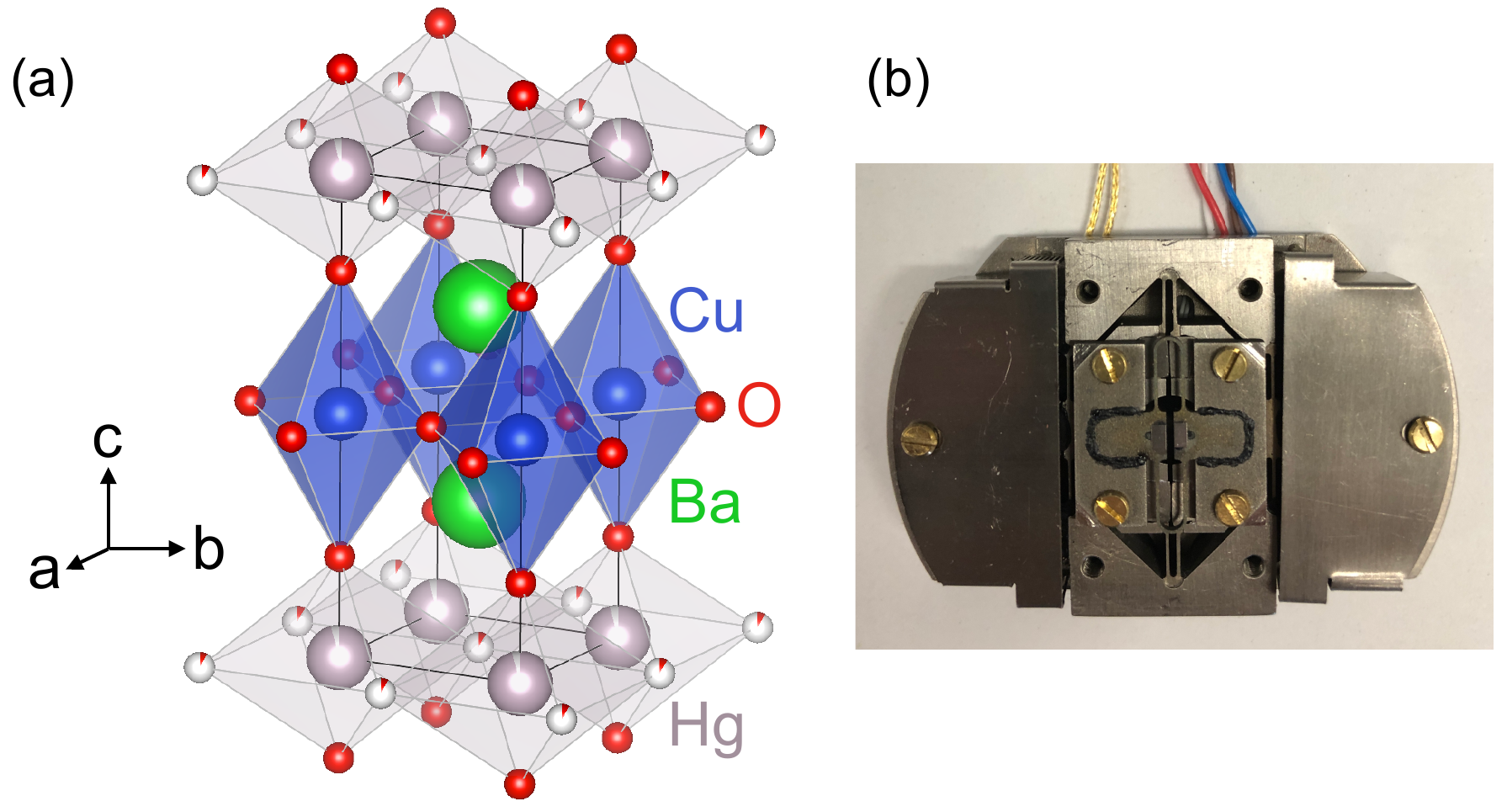}
\caption{\label{fig:Structure} (a) Crystal structure of HgBa$_2$CuO$_{4+\delta}$. The doped oxygen atoms stay in the Hg layers. (b) The Razorbill CS200T strain cell used to apply strain to the sample.}
\end{figure}

A natural choice for such exploration is the cuprate system HgBa$_2$CuO$_{4+\delta}$ (Hg1201), which has tetragonal symmetry, a single copper-oxygen layer, and minimal disorder \cite{Growth2008}.
The crystal structure of Hg1201, whose space group is $P4/mmm$ (No.123), is shown in Fig.~\ref{fig:Structure} (a). The absence of structural anisotropy due to its tetragonal symmetry implies that the response of Hg1201 to directional strain may differ fundamentally from that of orthorhombic YBCO. Strain-induced symmetry breaking which lifts the four-fold electronic degeneracies could, in principle, stabilize new ordering patterns. Moreover, the absence of bilayer splitting and chain contributions makes Hg1201 an ideal platform for studying the intrinsic properties of the CuO$_2$ plane. Previous X-ray scattering studies have identified short-range 2D CDW order with a wavevector \textit{H}\,=\,0.27 r.l.u. in samples with $T_c$\,=\,79\,K~\cite{Hg2017}. As the doping level is reduced and superconductivity weakens (e.g., $T_c$\,=\,55\,K), the CDW wavevector increases to \textit{H}\,=\,0.29 r.l.u.~\cite{Hg2017, Yu_PRX2020}, suggesting a close interplay between charge order and carrier concentration. Nevertheless, the CDW correlation length in Hg1201 remains very short: typically no more than a few lattice spacings—and no evidence for any long-range 3D CDW order has been reported to date. Whether an in-plane strain field can reinforce this fragile CDW order, modify its wavevector, or even induce a qualitatively new electronic phase, such as 3D CDW or electronic nematicity~\cite{Murayama_Natcom2019}, is an unresolved theoretical and experimental issue.

In this work, we investigate the structural changes and diffuse scattering features induced by \textit{a}-axis compressive strain in HgBa$_2$CuO$_{4+\delta}$ with $T_c$\,=\,78\,K. Our results reveal that the strain dependence of the lattice parameters in Hg1201 deviates significantly from that observed in YBCO, reflecting the different crystallographic and bonding environment of the Hg-based cuprate. Strikingly, we observe a strain-induced diffuse scattering signal with a wavevector near \textit{H}\,=\,0.5 r.l.u., which is not only distinct from the weak 2D CDW signal typically seen in Hg1201 but also suggests a new form of structural or electronic modulation. This diffuse scattering exhibits saturation at 0.2\% strain and shows little temperature dependence, hinting at a strain-stabilized, potentially static order. 

\section{Experimental and Computational Details\label{sec:Exp}}

\subsection{Sample preparation}

Single crystals of HgBa$_2$CuO$_{4+\delta}$ were grown using a self-flux method~\cite{Growth2006}. These crystals were annealed after growth over extended periods of time in air at 650\,$^\circ$C to achieve homogeneous doping. Sample quality was examined by X-ray single-crystal diffraction using a Rigaku MiniFlex 600 system and Laue diffraction by Photonic Science system. The superconducting transition temperature $T_c$ is determined to be 78$\pm$2\,K by the measurement of magnetic susceptibility using Quantum Design MPMS VSM equipment. In the following, the notation $T_c$ always means the transition temperature under zero uniaxial pressure.

The samples were cut using a wire saw and polished into a thin bar with dimensions 1.2\,mm along the $a$ axis, 0.2\,mm along the $c$ axis, and 75\,$\mu$m along the $b$ axis. The polished samples were mounted to the Razorbill CS200T strain cell by Loctite Stycast 2850FT epoxy with CAT 24LV as the epoxy catalyst [Fig.~\ref{fig:Structure} (b)]. We verified that the mounting process does not introduce additional strain by comparing the Raman spectra of unmounted and mounted samples.

\subsection{X-ray diffraction and thermal diffuse scattering}

The X-ray diffraction experiments were performed at the European Synchrotron Radiation Facility (ESRF) ID15B beamline. The incident beam with photon energy of 30.0\,keV was focused to a spot of 4\,$\mu$m diameter, and the diffracted beam was detected with a DECTRIS EIGER2 X 9M CdTe flat panel detector. Each diffraction image was recorded with an angular range of $\pm$35 degrees and an angular step of 0.5 degrees. Low-flux data were taken for structure refinement, and high-flux data were taken to better observe the diffuse scattering. The measured XRD data are available from the ESRF Data Portal~\cite{ESRF}.

The software CrysAlisPro by Rigaku Oxford Diffraction was used for cell refinement and data reduction. The programs SHELXL~\cite{SHELXL} and JANA~\cite{JANA} were used to solve the crystal structure and perform relevant refinements~\cite{SM}. For each refinement, around 500 Bragg peaks were used. The data were corrected for Lorentz, polarization, extinction, and absorption effects. The automatic unit cell finding rendered a primitive tetragonal lattice when the strain was zero, and a primitive orthorhombic lattice when the sample was compressed. The data reduction results were inspected by the corresponding scale factor vs. frame plot. The strain values reported in this work are the real strain determined from the measured lattice parameters.

\subsection{Raman scattering}

The Raman measurements were performed with a Horiba LabRAM HR Evolution spectrometer. A He-Ne laser (632.8\,nm) with less than 1\,mW power was focused to a spot of 5\,$\mu$m diameter with a x50 magnification objective. The spectra were collected with 1800\,mm$^{-1}$ (0.6\,cm$^{-1}$ spectral resolution) gratings and a liquid-nitrogen-cooled CCD detector. All the spectra were corrected for the instrumental spectral response and Bose factor. We used the $aa$ polarization configuration in which the incident light polarization is along the crystallographic $a$ direction, and the scattered one is also along the $a$ direction.

\subsection{First-Principles Lattice-Dynamics Calculations}
In order to interpret the X-ray scattering intensity between the Bragg reflections, simulations of the diffuse scattering induced by the thermal population of phonons were performed in first-order approximation assuming the validity of both harmonic and adiabatic approximations. The details of the formalism can be found in Ref.~\cite{TDS}.

The relevant dynamical matrices, necessary for the calculation of the thermal diffuse scattering, were calculated using the
linear response of density-functional perturbation theory as implemented
in the mixed-basis pseudopotential method \cite{Meyer,Heid1999}. In this approach
valence states are expanded in a basis set consisting of a combination of
plane waves and local functions, which allows an efficient description of
more localized valence states.
We used norm-conserving pseudopotentials including semi-core states Hg-5$p$, Hg-5$d$, Ba-5$s$, Ba-5$p$, Cu-3$s$, Cu-3$p$, and O-2$s$. The kinetic energy cutoff for the plane waves was 24 Ry, augmented by local functions of $s$,$p$,$d$ type at the Ba and Cu sites,
of $p$ and $d$ type at the Hg sites, and of $s$ and $p$ type at the O sites.
Brillouin zone integrations were performed with
a 8$\times$8$\times$4 tetragonal $k$-point grid using a Gaussian smearing of 0.2 eV.
We employed the PBESOL variant of the generalized gradient approximation
for the exchange-correlation functional \cite{Perdew2008}.
Dynamical matrices were calculated on a 4$\times$4$\times$2 momentum grid for both unstrained and strained HgBa$_2$CuO$_4$ using the experimental lattice constants~\cite{SM} and relaxing the internal structural parameters until the forces on the atoms were smaller than $2.6\times 10^{-2}$~eV/\AA.

\section{Results and Discussion\label{sec:Res}}

\subsection{Structural properties}

\begin{figure}
\includegraphics[width=0.98\linewidth]{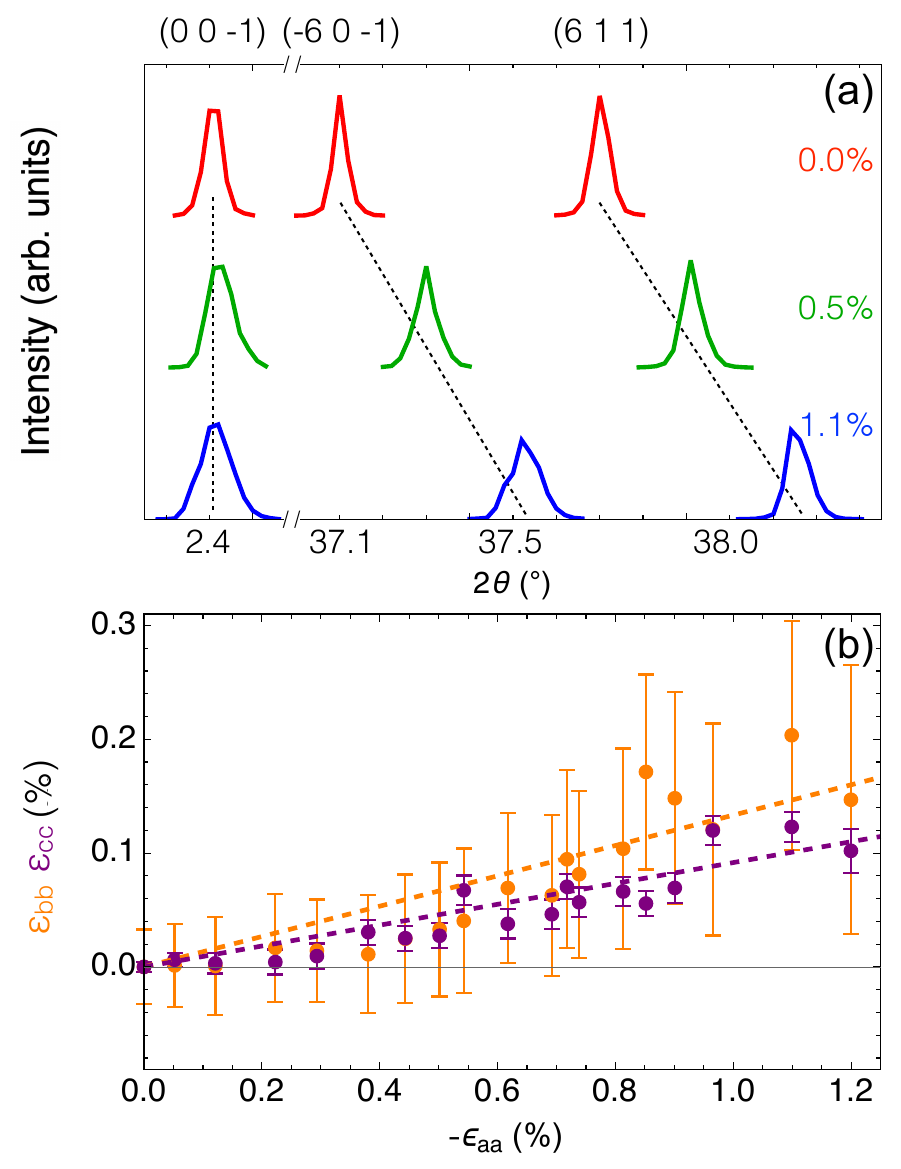}
\caption{\label{fig:Lattice} The strain-induced structural change in HgBa$_2$CuO$_{4+\delta}$ at $T_c$\,=\,78\,K. (a) The shift of three representative Bragg reflections as a function of the \textit{a}-axis compressive strain. The red, green, and blue colors represent 0.0\%, 0.5\%, and 1.1\% strain, respectively. The dashed black lines tracks the shift of each Bragg reflection under strain. The intensity of the Bragg peaks is normalized with respect to the (0\,0\,-1) Bragg peak at zero strain. (b) The magnitude of expansive strain in the \textit{b} ($\epsilon_{bb}$) and \textit{c} ($\epsilon_{cc}$) directions as a function of the magnitude of compressive strain in the \textit{a} direction ($-\epsilon_{aa}$). The strain values are determined from lattice parameters. The dashed lines are linear fits.}
\end{figure}

In Fig.~\ref{fig:Lattice} (a) we illustrate how the applied compressive strain leads to a structural change by presenting the shift of three Bragg reflections. Because the structural change mainly happens along the $a$ axis, the shift is mostly dependent on the H value of the Bragg reflections: the $2\theta$ value of both (6\,1\,1) and (-6\,0\,1) reflections increases with strain, whereas that of the (0\,0\,-1) reflection exhibits no shift instead. Another feature is that the linewidth of the Bragg reflections broadens with strain, probably due to strain-induced disorder or inhomogeneity. The strain-induced structural change in the $b$ and $c$ directions is shown in Fig.~\ref{fig:Lattice} (b). Noticeably, the change of the lattice parameters along the $b$ and $c$ axes stays within 0.2\% up to 1.2\% strain, showing that these two lattice parameters respond weakly to the compression along the $a$ axis. By doing a linear fit to the relative change of the lattice parameters, the Poisson ratios are estimated to be $\nu_{ba}$\,=\,0.16 and $\nu_{ca}$\,=\,0.11, which are reasonable values for ceramics~\cite{Review2011}. For comparison, the Poisson ratios in YBCO are $\nu_{ba}$\,$\sim$\,0.4 and $\nu_{ca}$\,$\sim$\,0.15~\cite{Y2024}.

The Cu-O distances are a crucial factor influencing the superconducting transition temperature~\cite{Pavarini_PRL2001, Peng_2017}. In both $a$ and $b$ directions, the Cu-O bond length and the lattice parameter have the same percentage changes~\cite{SM}, because of the structural simplicity of the Cu-O plane. However, the out-of-plane Cu-O distance increases by 0.86\% whereas the \textit{c}-axis lattice parameter increases only by 0.12\% under 1.1\% strain~\cite{SM}. For comparison, in-plane uniaxial compression leads to almost no change of the apical oxygen position with respect to the CuO$_2$ plane in YBCO~\cite{Y2024}. Our structural analysis, therefore, reveals that in-plane uniaxial compression brings the apical oxygen away from the CuO$_2$ planes, a behavior typically associated with an increase in the superconducting transition temperature~\cite{Pavarini_PRL2001, Peng_2017}.

\subsection{Diffuse scattering}
\begin{figure}
\includegraphics[width=0.98\linewidth]{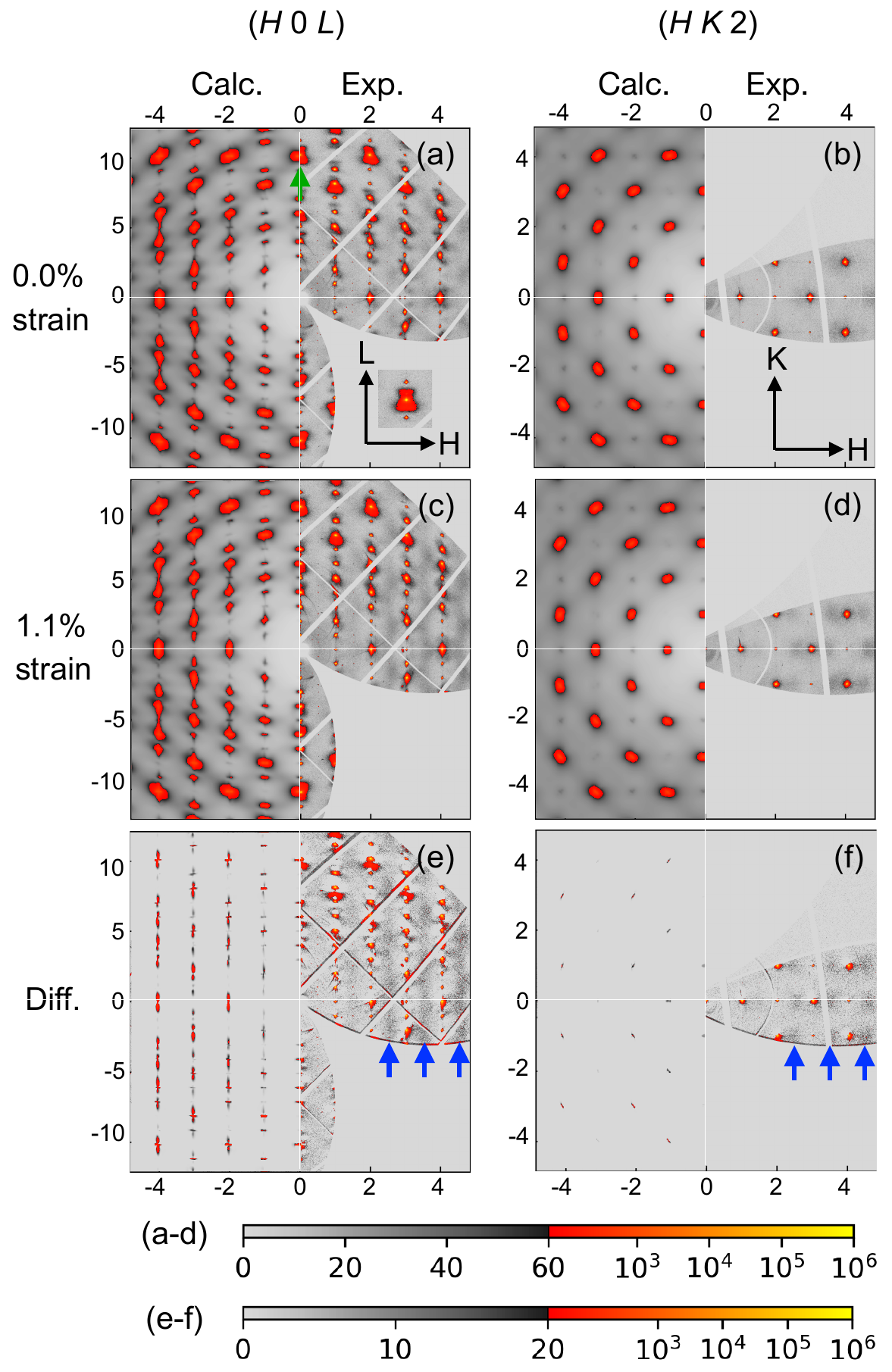}
\caption{\label{fig:Diff} The experimental X-ray diffraction patterns measured at $T_c$\,=\,78\,K (shown in the right side of each panel, Exp.) compared with the calculated thermal diffuse scattering (shown in the left side of each panel, Calc.) for HgBa$_2$CuO$_{4+\delta}$. The three rows from top to bottom shows the results at zero strain, 1.1\% strain, and their difference, respectively. The three panels to the left are in the (\textit{H}\,0\,\textit{L}) plane, whereas the three panels to the right are in the (\textit{H}\,\textit{K}\,2) plane. The strain-induced diffuse scattering, which has strip shape in panel (e) and dot shape in panel (f), is labeled by blue arrows. The bow-tie-shaped lobes along the L direction is labeled by a green arrow and enlarged at bottom-right corner in panel (a).}
\end{figure}

In Fig.~\ref{fig:Diff} we present a comprehensive survey of the strain-dependence of the experimentally measured X-ray scattering intensity across selected reciprocal space planes at the superconducting transition temperature $T_c$. This temperature was chosen based on prior studies of YBCO, which reported the most pronounced strain effects occurring in the vicinity of $T_c$~\cite{Y2024, Y2018}. 
Our analysis primarily focuses on the diffuse scattering intensity between the main structural Bragg reflections, where weak signatures of short-range CDW modulations are expected.

The X-ray scattering intensity in the reciprocal (\textit{H}\,0\,\textit{L}) and 
(\textit{H}\,\textit{K}\,2) planes in the absence of applied strain is shown in Fig.~\ref{fig:Diff} (a) and (b), respectively.
The Bragg reflections are resolution-limited and intense compared to the diffuse background. The broad base of Bragg reflections, especially in Fig.~\ref{fig:Diff} (a), appears elongated along specific high-symmetry directions.
This broadening arises from the thermal diffuse scattering of acoustical phonons, and can be well captured by the calculated diffuse scattering.

Moreover, the experimental results show broad, bow-tie-shaped lobes existing along \textit{L} direction from some Bragg reflections with large \textit{L} values [Fig.~\ref{fig:Diff} (a), identified by the green arrow and enlarged at bottom-right corner]. These features have previously been reported and associated with nanoscale correlations of atomic displacements perpendicular to the CuO$_2$ planes~\cite{Greven2024,Review2025}, but cannot be easily reproduced by our calculations.

The experimentally accessible reciprocal space in the (\textit{H}\textit{K}2) plane [Fig.~\ref{fig:Diff} (b)] is more limited due to the scattering geometry.
Because of the structure factors, reflections at \textit{H}+\textit{K}\,=\,2\textit{n} are much weaker than those at \textit{H}+\textit{K}\,=\,2\textit{n}\,+\,1.
The reflection at (1\,0\,2) appears sharper than other reflections with larger \textit{H}+\textit{K}\,=\,2\textit{n}\,+\,1 values [for example, (4\,1\,2)], consistent with the calculation results.

We note that the sharp diffuse lines observed in optimally doped Hg1201~\cite{OD2011} and associated with the formation of dopant O-chains are not seen in our diffraction patterns, as expected for underdoped materials. Moreover, we do not observe here the weak signature of 2D CDW order that has been reported at in-plane wavevector \textit{H}\,=\,0.27 r.l.u. in a series of resonant scattering experiments~\cite{Tabis_NatCom2014, Yu_PRX2020, Hg2017, Raman2020}.

Applying compressive 1.1\% strain along the crystallographic \textit{a} direction leads to additional diffuse scattering intensity between the reflections. 
In the (\textit{H}\,0\,\textit{L}) plane, this strain-induced diffuse scattering forms strips at half-\textit{H} values along the \textit{L} direction [Fig.~\ref{fig:Diff} (c)], showing that the corresponding modulation is in-plane confined. 
In the (\textit{H}\,\textit{K}\,2) plane, the same diffuse-scattering features manifest themselves as broad peaks at the two sides of Bragg peaks along the \textit{H} direction [Fig.~\ref{fig:Diff} (d)]. 
To better emphasize the strain-induced contribution to the diffuse scattering, we subtract the zero-strain diffraction patterns [Fig.~\ref{fig:Diff} (a) and (b)] from the 1.1\%-strain diffraction patterns [Fig.~\ref{fig:Diff} (c) and (d)]. Fig.~\ref{fig:Diff} (e) and (f) unambiguously indicate that the strain-induced diffuse scattering has a rod-like structure along the \textit{L} direction and is centered in-plane around wavevectors with \textit{H}\,=\,0.5 r.l.u. 
Most interestingly, this experimentally observed strain-induced change cannot be simply explained by the change of the calculated phonon dispersion.
As seen in Fig.~\ref{fig:Diff} (e) and (f), the impact of strain on the calculated thermal diffuse scattering is strictly limited to regions close to the Bragg reflections and relates to the directional stiffening of the lattice under uniaxial compression.

\begin{figure}
\includegraphics[width=0.98\linewidth]{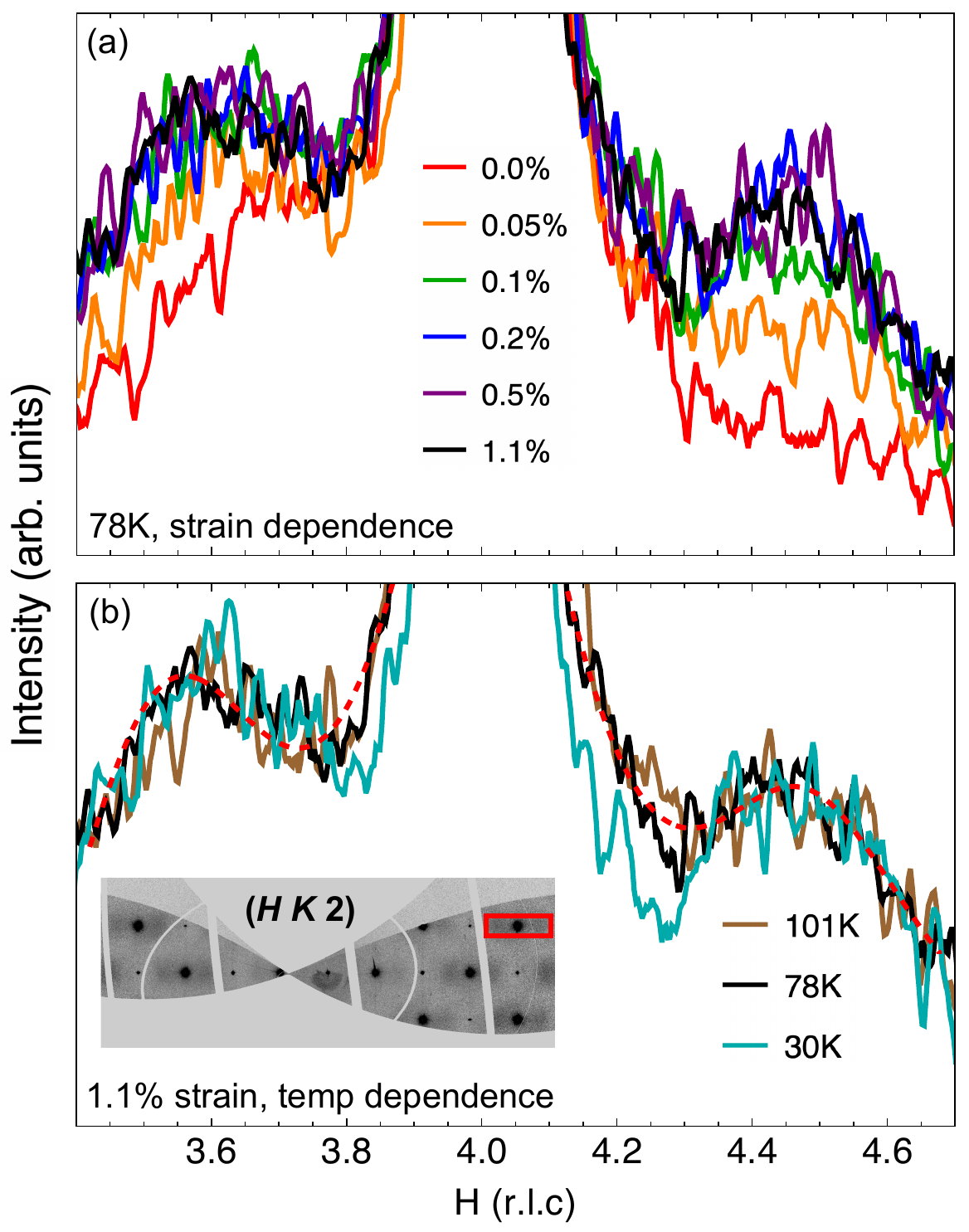}
\caption{\label{fig:Cut} The line cuts along H direction across the (4\,1\,2) Bragg peak. (a) The strain dependence of the diffuse scattering at $T_c$\,=\,78\,K. (b) The temperature dependence of the diffuse scattering under 1.1\% strain. The dashed red curve show the fit to the 78\,K spectrum. The (4\,1\,2) Bragg peak and the diffuse scattering associated with it are labeled by an red rectangle in the diffraction pattern shown at bottom-left corner.}
\end{figure}

Next, to gain more insights regarding the nature of the strain-induced signal, we examine in Fig.~\ref{fig:Cut} its strain and temperature dependence. To achieve this, cuts are made across these features along the reciprocal \textit{H} direction. These cuts reveal that the signal already appears with minimal compression of 0.05\% and rapidly saturates for in-plane strain as low as 0.2\% [Fig.~\ref{fig:Cut} (a)]. Furthermore, at 1.1\% compressive strain, there is no significant impact of temperature on these features. Moreover, we do not observe any particular loss of intensity in the superconducting state (at 30\,K) or the normal state (at 101\,K) [Fig.~\ref{fig:Cut} (b)].

Finally, we analyse quantitatively the lineshape of the strain-induced features by fitting them to a Lorentzian profile in the \textit{H} direction: $I(q) \propto 1/[\left(q-q_{0}\right)^{2}+\kappa^{2}]$, in which $q_{0}$ is the center position and $\kappa$ is the decay factor. The Lorentzian lineshape corresponds in real space to a short-range order whose correlation function follows exponential decay: $G(x) \propto e^{-x / \xi}$, in which the correlation length $\xi=1/\kappa$. By fitting the strain-induced diffuse scattering, we find the correlation length to be around 4 lattice constants. The spectral linewidth and, in turn, the correlation length show little temperature or strain dependence. The $q_{0}$ values for the two strain-induced features shown in Fig.~\ref{fig:Cut} (b) are very close to half integer values, 3.53(4) and 4.48(2) r.l.u., respectively, consistent with a nearly-commensurate nature.

\subsection{Raman scattering}
\begin{figure}
\includegraphics[width=0.98\linewidth]{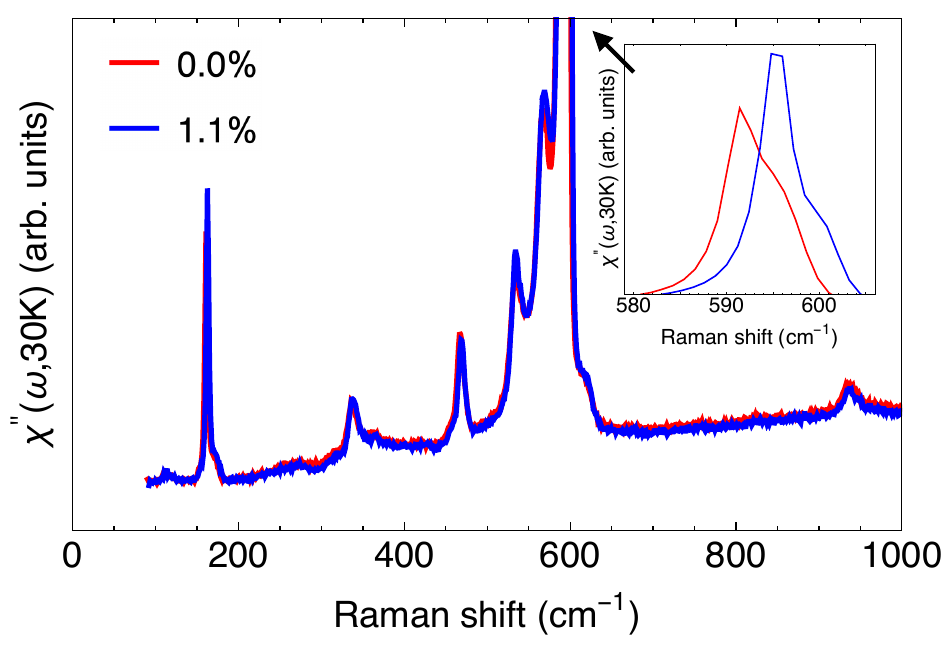}
\caption{\label{fig:Raman} Strain dependence of Raman spectra measured in the \textit{aa} polarization geometry at 30\,K. The inset shows the shift of the A$_{1g}$-symmetry phonon at 592\,cm$^{-1}$.}
\end{figure}

To determine whether the strain-induced signal may be related to strain-induced changes in short-range oxygen dopant structures, we performed a Raman scattering study.
We show in Fig.~\ref{fig:Raman} the Raman spectra measured in the \textit{aa} polarization geometry at 30\,K. The compound HgBa$_2$CuO$_{4+\delta}$ has four Raman-active phonon modes: 2A$_{1g}$ and 2E$_{g}$ which have previously been investigated~\cite{Raman1996, Zhou_PRB96}. According to Raman selection rules~\cite{Hayes1978}, Raman signal with A$_{1g}$ and B$_{1g}$ symmetry is probed in the \textit{aa} polarization geometry. In Fig.~\ref{fig:Raman}, the modes at 161 and 592\,cm$^{-1}$ correspond to the A$_{1g}$ phonon derived from Ba and O atoms, respectively~\cite{Raman1996}; the other relatively weak modes have been assigned to defects arising from oxygen dopants~\cite{Raman1996,Raman2020, Zhou_PRB96}. 
Fig.~\ref{fig:Raman} shows that \textit{a}-axis compressive strain has only moderate impact on the zone center phonons of Hg1201. When the strain is increased to 1.1\%, the frequency shift for the 161\,cm$^{-1}$ and 592\,cm$^{-1}$ A$_{1g}$-symmetry phonons is around 1\,cm$^{-1}$ and 4\,cm$^{-1}$, respectively. The corresponding Gr\"uneisen parameters, $\gamma_i = -(\Delta \omega_i/\omega_i)/(\Delta V/V)$, for these two phonons at 30\,K are 0.8 (the 161\,cm$^{-1}$ mode) and 0.6 (the 592\,cm$^{-1}$ mode), respectively. For a comparison, the Gr\"uneisen parameter for the 592\,cm$^{-1}$ A$_{1g}$-symmetry phonon derived from measurements under hydrostatic pressure is 0.52~\cite{G1995} or 0.61~\cite{G2021,Tc2014} (We note that the Gr\"uneisen parameters obtained from uniaxial compression and hydrostatic pressure experiments do not need to be the same.)
Moreover, we do not observe any significant strain-induced change in the defect modes, nor do we detect the appearance of any new features, confirming the absence of noticeable strain-induced structural distortions.

\subsection{Discussion}
We have demonstrated that the application of uniaxial pressure along the crystallographic \textit{a}-axis of underdoped HgBa$_2$CuO$_{4+\delta}$ induces a previously unreported modulation characterized by several intriguing features.

First, the strain-induced modulation manifests as a pronounced rod-like structure in the (\textit{H}\,0\,\textit{L}) plane, clearly indicative of the absence of correlations along the c-axis and reinforcing the two-dimensional nature of this phenomenon. Within the CuO$_2$ plane, satellite peaks exclusively appear along the \textit{H} direction, suggesting an underlying unidirectional structure. Furthermore, the strain-induced signal saturates already at relatively modest strain levels (0.2\%) and, notably, is only detected along the direction parallel to the applied strain.

Second, the observed modulation emerges at a nearly-commensurate wave vector corresponding to a doubling of the unit cell. This behavior sharply contrasts with previous observations in underdoped cuprates, where charge density waves typically manifest at incommensurate wave vectors associated with periodicities spanning approximately 3 to 4 unit cells~\cite{RXS2016}. Remarkably, Hg1201 is thus far the only known cuprate exhibiting commensurate low-energy magnetic fluctuations, as previously reported in Refs.~\cite{Yuan2010, Yuan2010PRB, Greven2016a, Greven2016b}. Moreover, the observed modulation displays an in-plane correlation length similar to these magnetic fluctuations, suggesting a potentially unique and intimate interplay between charge and spin correlations specific to this compound.

Third, the absence of a significant temperature dependence of the strain-induced signal indicates insensitivity to the onset of superconductivity, consistent with earlier reports of charge ordering phenomena in this family of materials~\cite{Tabis_NatCom2014, Hg2017, Yu_PRX2020}. However, further studies will be necessary to clarify this relationship, along with investigations into the effect of uniaxial pressure on the superconducting properties of Hg1201, which remain largely unexplored.

We conclude this discussion by emphasizing that although this particular type of nearly-commensurate charge correlation has yet to be observed in other cuprate families, it bears a striking resemblance to degenerate site-charge density wave stripes characterized by wave vectors $(\pi, 0)$ and $(0, \pi)$. Interestingly, these stripe configurations emerge naturally within the mean-field phase diagram derived from the $\pi$-flux state employing fermionic spinon approaches to the resonating valence bond spin-liquid model on the square lattice~\cite{Subir2003, Christos_PNAS2023, Sachdev2025}. This analogy is particularly compelling given that Hg1201 is among the very few cuprates possessing a simple tetragonal crystal structure and minimal doping-induced disorder. These unique attributes establish Hg1201 as an ideal candidate for unraveling the intrinsic electronic properties inherent to doped CuO$_2$ planes.

\section{Conclusion\label{sec:Con}}

We use synchrotron X-ray diffraction to study the structural change and diffuse scattering induced by \textit{a}-axis compressive strain in HgBa$_2$CuO$_{4+\delta}$ with $T_c$\,=\,78\,K. The lattice parameters in the \textit{b} and \textit{c} directions respond weakly to the compression in the \textit{a} direction, which leads to relatively small Poisson ratios of $\nu_{ba}$\,=\,0.16 and $\nu_{ca}$\,=\,0.11. Regarding the Cu-O distances, although the \textit{c}-axis lattice parameter increases by only 0.1\% at 1.1\% \textit{a}-axis strain, the Cu-O distance along the \textit{c} axis increases by 0.9\%. By comparing the experimental diffraction patterns with the simulation of thermal diffuse scattering, we identify strain-induced diffuse scattering with a wavevector near (0.5\,0\,0) and a correlation length of around 4 unit cells along the \textit{a} direction. The diffuse-scattering features saturate at 0.2\% strain and exhibit little change on cooling below $T_c$. We suggest that such diffuse scattering corresponds to a new 2D charge correlation which does not compete with the superconductivity. The nearly-commensurate wavevector of this new charge modulation bears a resemblance to the magnetic fluctuations in this system, and such order was found to emerge in the mean-field phase diagram derived from the $\pi$-flux state employing fermionic spinon approaches to the resonating valence bond spin-liquid model on the square lattice. Our findings provide new insights into the coupling between strain and charge order in structurally simple cuprates and open new directions for engineering electronic phases in high-$T_c$ superconductors via lattice perturbations.

\begin{acknowledgments}

We acknowledge Steve Kivelson and Subir Sachdev for fruitful discussion.
The work at Karlsruhe Institute for Technology was funded by the Deutsche Forschungsgemeinschaft (DFG, German Research Foundation) - TRR 288 - 422213477 (project B03), and Projektnummer 449386310.
The work at Peking University was supported by the National Natural Science Foundation of China (Grant No. 12061131004) and by the National Basic Research Program of China (Grant No. 2021YFA1401901).
M. F. acknowledges funding from the Alexander von Humboldt foundation, and the YIG preparation program of the Karlsruhe Institute of Technology.
R.H. acknowledges support by the state of Baden-W\"{u}rttemberg through bwHPC.

\end{acknowledgments}


%

\end{document}